\documentclass{article}
\usepackage[a4paper,total={6.6in,10in}]{geometry}
\usepackage{amsmath}  
\usepackage{amsfonts} 
\usepackage{graphicx} 
\usepackage{xcolor}
\usepackage{endnotes}
\usepackage{soul}

\usepackage[colorlinks = true,
            linkcolor = blue,
            urlcolor  = blue,
            citecolor = blue,
            anchorcolor = blue]{hyperref}
\usepackage{subfig}
\begin{document}
\title{Computing Scattering Cross Sections For Spherically Symmetric Potentials}
\author{Anil Khachi$^{1}$\\\\
$^{1}$Department of Physics,\\ St. Bede's College, 171002, \\Himachal Pradesh, India\\\\}
\date{}
\maketitle
\noindent \textbf{Note:} This paper is accepted in American Journal of Physics (26/04/2024)\\
\\
\noindent This paper introduces students and instructors to the use of Scilab for calculating phase shifts from phenomenological potentials in nuclear, atomic, and molecular scattering, beginning with a Riccati-type nonlinear differential equation derived from the time-independent Schr$\ddot{\text{o}}$dinger equation. For spherically symmetric potentials, this equation can be easily solved using Xcos, a Scilab toolbox. \cite{PRC}\cite{bobby} Scilab is open source software that is used for numerical computations and can be freely downloaded for Windows, Linux and Mac OS.\cite{scilab} Xcos is a Scilab toolbox dedicated to the modeling and simulation of hybrid dynamic systems \cite{rachna}. Xcos provides users with the ability to model complex dynamical systems using a block diagram editor (GUI based), offering a step-by-step procedure for learning to solve complex differential equations interactively. 

The normal method for calculating scattering cross sections involves solving the TISE in the interior ($V \neq 0$) and exterior regions and matching boundary conditions to find the partial wave amplitudes, $C_{\ell,m}$. \cite{Erich}\cite{griffiths}
 
The Numerov method can also be used to obtain the phase shift \cite{Erich}, but it requires knowledge of the wavefunction. In contrast, using the Variable Phase Approach (VPA), it is not necessary to first calculate radial wave functions and then to find the phases according to their asymptotic behaviour. The VPA is a valuable teaching tool because it shows how the scattering phase shift is affected by the presence of a potential. In Calogero's words: ``\textit{It is important to acquire a physical feeling for the relation between the potential and the phase shift; this feeling, we believe, is provided by the variable phase approach}". \cite{Calogero}

The VPA was originally proposed by Morse and Allis \cite{Morse}, and later came to be known as the phase function method \cite{Calogero}. The VPA has been applied to understand various nuclear scattering reactions like np and $\alpha-\alpha$ \cite{PRC}\cite{scripta}. In this approach the TISE for a particle with wavenumber \textit{k} and orbital angular momentum $\ell$ undergoing scattering is transformed into a first order non-linear Ricatti-type equation.

The TISE for a particle with energy \textit{E}, reduced mass $\mu$, and orbital angular momentum $\ell$ undergoing scattering by a potential \textit{V}(r) is given by
\begin{equation}
\frac{\hbar^2}{2\mu} \bigg[\frac{d^2}{dr^2}+\bigg(k^2-\frac{\ell(\ell+1)}{r^2}\bigg)\bigg]u_{\ell}(k,r)=V(r)u_{\ell}(k,r).
\label{Scheq}
\end{equation}
This linear homogeneous second order differential equation can be transformed into Riccati type phase function equation \cite{Calogero}, containing phase shift information with initial condition $\delta_\ell(k,0)=0$, given by 

\begin{equation}
\small{
\frac{d\delta(k,r)}{dr}=-\frac{U(r)}{k}\big[\cos\delta_\ell(k, r)\hat{j}_{\ell}(kr)-\sin\delta_\ell(k, r)\hat{\eta}_{\ell}(kr)\big]^2
\label{PFMeqn}}
\end{equation}
where $U(r)=V(r)/(\hbar^2/2\mu)$, $\hat{j_{\ell}}(kr)$ and $\hat{\eta_{\ell}}(kr)$ are Riccati-Bessel and Riccati-Neumann functions respectively. 

The dimensionless phase shift $\delta(k)$ is obtained from the phase function $\delta(k, r)$ through the limiting process:
$\delta(k)=\lim_{r \to \infty} \delta_\ell(k, r)$. 
It should be noted that the VPA applies only to a restricted class of potentials with a definite behaviour for $r\rightarrow\infty$. If the potential does not become negligible in the asymptotic region, then the phase function equation becomes ill-defined. For example, in interactions involving charged systems such as proton-proton or $\alpha-\alpha$ interactions, the Coulomb potential does not diminish rapidly to zero; more detailed analysis has been provided by Laha \textit{et al.} \cite{lahaji}. For such cases, a screened Coulomb potential, which behaves like a Coulomb potential for smaller \textit{r} and decreases exponentially for large \textit{r}, such as the Hult\'{e}n potential \cite{lahaji}, can be utilised for the phase function equation to be valid \cite{fuji}. Additionally, it should be noted that obtaining a sufficiently accurate numerical solution of Eq. \ref{PFMeqn} requires much more care than finding a very accurate numerical solution of Eq. \ref{Scheq}, as the latter equation is nonlinear. The VPA also encounters limitations when computing phase shifts for higher partial waves (typically $\ell >$ 6-8) as the Bessel functions involves more intricate mathematical expressions.
\begin{figure}
    \centering
    \includegraphics[width=0.8\textwidth, height=11cm]{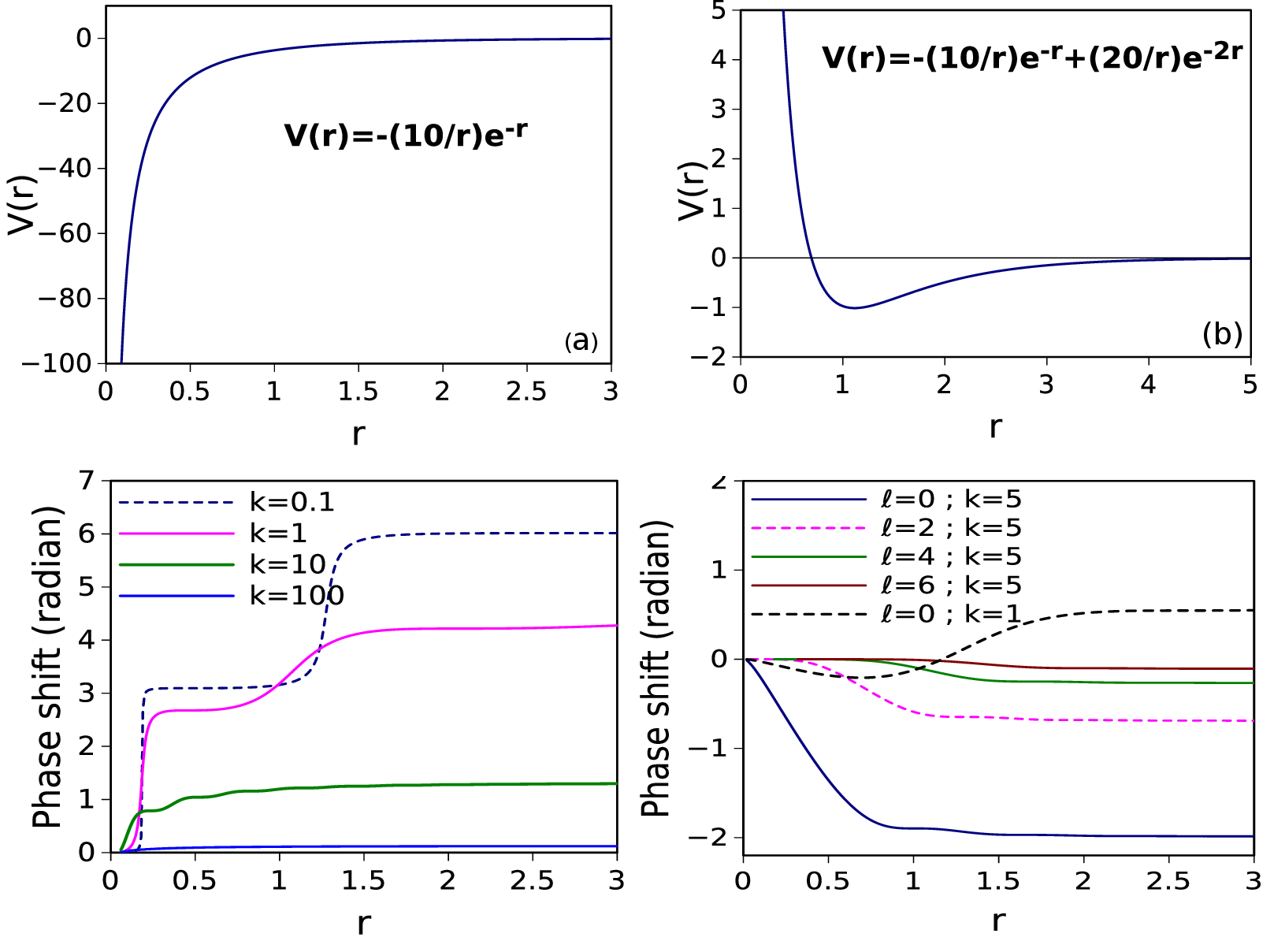} 
    \caption{The upper plots show (a) the attractive  Yukawa potential and (b) the attractive+repulsive Yukawa potential. The lower plots show the resulting phase shifts for a range of \textit{k}' s}
    \label{fig:att_pot_l2}
\end{figure}
Advantages of VPA/PFM over traditional methods of solving TISE are: 
\begin{enumerate}
\item The phase function is sufficient for the complete determination of the wave function
\item At each part of potential we can see how the scattering phase develops in space
\end{enumerate}
More advantages of VPA have been discussed by Babikov \cite{babikov}. Figure \ref{fig:att_pot_l2} shows the phase shift calculation for two spherically symmetric potentials: the attractive Yukawa potential and a combination of attractive and repulsive Yukawa interactions. The Xcos code that generated these plots is remarkably simple and is shown in the supplementary material \cite{supp}. One can further obtain the scattering length 
\textit{a}, a parameter that provides an effective size of the potential and indicates how much the presence of the potential alters the trajectory at low energies, i.e., near \textit{k}=0, by simply using the formula
$k \cot \delta_0(k,r)\approx-1/a+0.5r_ek^2$ \cite{darewych}. 
More examples along with their Xcos models are given in the supplementary material \cite{supp}. 

In conclusion we quote Calogero: ``\textit{VPA is very helpful in most problems of current physical interest, including those in high-energy physics where a simple potential model is certainly inadequate to describe the dynamics completely}".

\section*{Acknowledgements}
This paper is dedicated to my father Mr. L. R. Khachi for his love and affection. I would also like to express my sincere gratitude to the anonymous reviewers for their invaluable feedback and constructive criticism which have greatly improved the quality of this paper.
\newpage
\section*{~~~~~~~~~~~~~~~~~~~~~~~~~~\textbf{Supplementary Material}}
\vspace{1cm}
\section{Detailed Methodology}
The Schr$\ddot{\text{o}}$dinger wave equation for a particle with energy \textit{E} and orbital angular momentum $\ell$ undergoing scattering is given by
\begin{equation}
\frac{\hbar^2}{2\mu} \bigg[\frac{d^2}{dr^2}+\bigg(k^2-\frac{\ell(\ell+1)}{r^2}\bigg)\bigg]u_{\ell}(k,r)=V(r)u_{\ell}(k,r)
\label{Scheq}
\end{equation}
where $k = \sqrt{E/(\hbar^2/2\mu)}$.
The second order differential equation Eq.\ref{Scheq} can been transformed into a pair of first order differential equations: one is a first order phase function equation i.e $\delta'_{\ell}$(k,r) (Eq.\ref{PFMeqn}) and another equation involves both amplitude function $A_{\ell}$(k,r) and the phase function $\delta_{\ell}$(k,r) called amplitude function $A'_{\ell}$(k,r) \cite{Calogero}\cite{babikov}. The phase function equation given by \cite{Calogero}
\begin{equation}
\delta_{\ell}'(k,r)=-\frac{2\mu}{\hbar^2}\frac{U(r)}{k}\bigg[\cos(\delta_\ell(k,r))\hat{j}_{\ell}(kr)-\sin(\delta_\ell(k,r))\hat{\eta}_{\ell}(kr)\bigg]^2
\label{PFMeqn}
\end{equation}
The phase shift $\delta(k)$ is obtained from the phase
function $\delta(k, r)$ through the limiting process:
\begin{equation}
\delta(k)=\lim_{r \to \infty} \delta_\ell(k, r)
\end{equation}
In integral form the above equation can be written as
\begin{equation}
\delta(k,r)=-\frac{2\mu}{\hbar^2}\frac{1}{k}\int_{0}^{r}{U(r)}\bigg[\cos(\delta_{\ell}(k,r))\hat{j_{\ell}}(kr)-\sin(\delta_{\ell}(k,r))\hat{\eta_{\ell}}(kr)\bigg]^2 dr
\end{equation}
Here $\hat{j}_{\ell}$ and $\hat{\eta}_{\ell}$ are Riccati-Bessel and Riccati-Neumann functions and $U(r)=V(r)/(\hbar^2/2\mu)$. Since we have not taken any physical scattering system in this paper so we have considered $(\hbar^2/2\mu)=1$. Normally $V(r)$ is measured in units of MeV, $\hbar^2/2\mu$ in MeV.fm$^{2}$ and wavenumber \textit{k} in fm$^{-1}$, as a result overall $\delta(k,r)$ comes out to be dimensionless quantity with unit of  radians or degrees. For $\ell=0$, $\hat{j}_{0}(kr)=\sin(kr)$ and $\hat{\eta}_{0}(kr)=-\cos(kr)$, hence equation \ref{PFMeqn} takes following simple form:
\begin{equation}
\frac{d\delta_0(r)}{dr} = -\frac{V(r)}{k}\sin^2(kr+\delta_0(r))
\label{pheqn}
\end{equation}
The value of $\delta_0$ at $r = 0$ is chosen to be zero, as interaction with potential is not present, since $u_0(r=0)=0$. This Eq. \ref{pheqn} can not be easily solved using analytical method and hence we take up numerical approach. 

For higher partial waves, the Riccati-Bessel and Riccati-Neumann functions used in VPA/PFM can be easily obtained by using following recurrence formulae:
\begin{equation*}
    \hat{j}_{\ell+1}(kr)=\frac{2\ell+1}{kr} \hat{j_\ell}(kr)-{\hat{j}_{\ell-1}}(kr)~~~~~~~~~~~~~~~~~~~~~~~~~~~~~~~~~~~~~~~~~~~~~~~~~~(A1)
    \label{R1}
\end{equation*}
\begin{equation*}
   \hat{\eta}_{\ell+1}(kr)=\frac{2\ell+1}{kr} 
   \hat{\eta_\ell}(kr)-{\hat{\eta}_{\ell-1}}(kr)~~~~~~~~~~~~~~~~~~~~~~~~~~~~~~~~~~~~~~~~~~~~~~~~~(A2)
   \label{R2}
\end{equation*}
For $\ell$=1 \& 2, phase function equation takes following form:
\begin{equation*}
\delta_1'(k,r)=-\frac{V(r)}{k}\bigg[\frac{\sin(\delta_1+(kr))-(kr) \cos(\delta_1+(kr))}{(kr)}\bigg]^2
\end{equation*}

\begin{equation*}
\delta_2'(k,r) = -\frac{V(r)}{k}\bigg[-\sin{\left(\delta_2+(kr) \right)}-\frac{3 \cos{\left(\delta_2 +(kr)\right)}}{(kr)} + \frac{3 \sin{\left(\delta_2 + (kr) \right)}}{(kr)^2}\bigg]^2 
\label{NS3}
\end{equation*}
Similarly, phase function equations can be obtained for higher partial waves using equations A1 and A2. Implementation of above VPA equations for all $\ell$ channels are carried out in Scilab's Xcos environment \cite{Scilab}
\section{Results and Discussions}
The scattering phase shifts have been obtained for following four spherically symmetric potentials:
\begin{enumerate}
\item Kronig-Penney type periodic potential
\item Attractive Yukawa
\item Repulsive Yukawa
\item Combination of attractive and repulsive Yukawa interaction
\end{enumerate}
\subsubsection{Numerical Implementation of Phase Equation}
\noindent To solve the phase equation for different $\ell$ channels, one need to specify the potential U(r) and energies involved through `k'. To test the algorithm and understand the nature of solutions, we have chosen the following three spherically symmetric potentials:
\begin{itemize}
\item \underline{Kronig-Penney type periodic potential:} The Xcos model for Kronig-Penney type periodic potential is shown in Fig.\ref{sqw}. The potential \textit{V}(r) is a periodic square wave as shown in Fig.\ref{penny}, and the corresponding phase shift is shown moving like stairs above the periodic potential. It can be clearly seen that the phase shift becomes constant when the potential is constant in accordance with VPA equation.
\begin{figure}[h!]
      \includegraphics[scale=0.5]{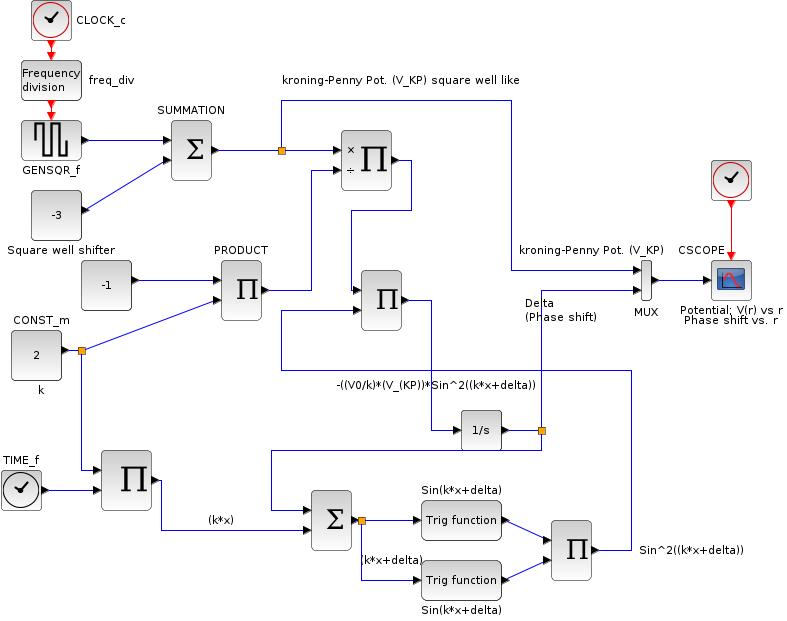}
    \caption{Scilab-Xcos scheme for phase shift variation for Kronig-Penney type periodic potential.}
    \label{sqw}
\end{figure}
\begin{figure}[h!]
\centering
      \includegraphics[scale=0.55]{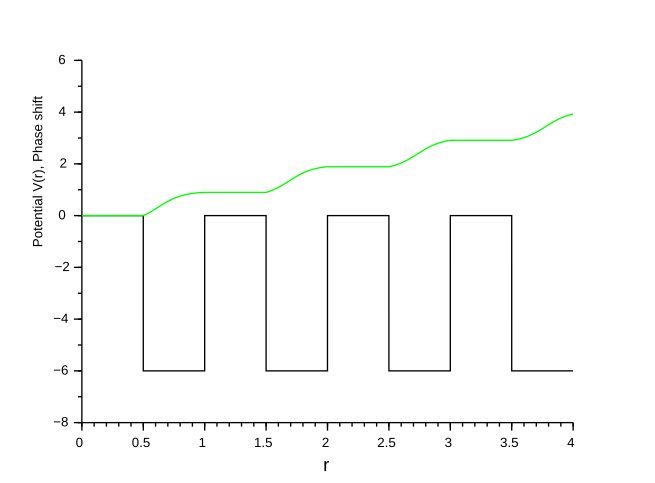}
    \caption{Kronig-Penney type periodic potential (black) and phase shift vs. distance variation (green) for Kronig-Penney type potential for k=2.}
    \label{penny}
\end{figure}
\begin{figure}[h!]
\centering
\hspace*{-3cm}
      \includegraphics[scale=0.45]{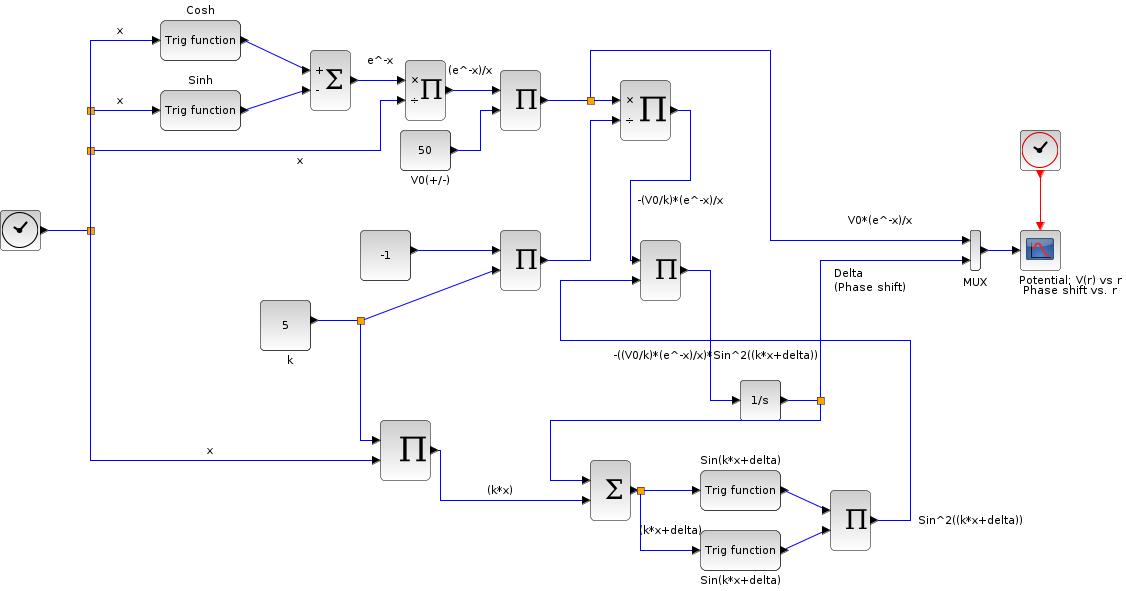}
    \caption{Scilab-Xcos scheme for phase shift variation for attractive ($V(r)=-V_0/re^{-r}$) or repulsive potential ($V(r)=+V_0/re^{-r}$) w.r.t distance.}
    \label{att_rep_scheme}
\end{figure}

\begin{figure}[h!]
\centering
\hspace*{-1.5cm}
{\includegraphics[scale=0.6]{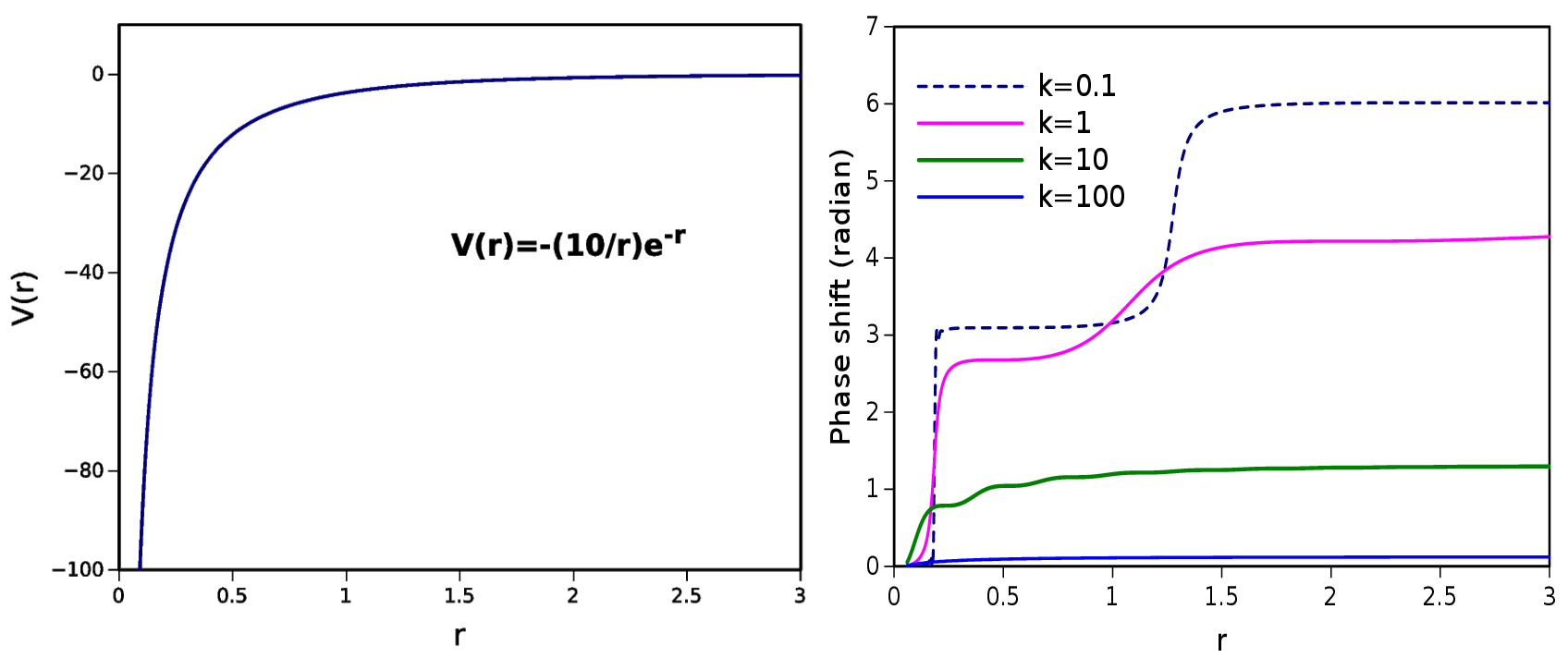}}
\caption{Phase shift variation with \textit{r} for  attractive Yukawa potential $V(r)=-10/re^{-r}$ for $\ell=0$.}
\label{att_rep_plot}
\end{figure} 
\item \underline{Attractive Yukawa:} The potential considered is having form $V(r)= -(10/r)e^{-r}$ and is being solved using VPA for $\ell$ = 0 and 2 partial waves. The Xcos model for attractive/repulsive Yukawa type potential \cite{Yukawa} is given in Fig.\ref{att_rep_scheme}. The $\ell=0$ phase shifts as a function of \textit{r} are plotted for values of \textit{k} = 0.1, 1, 10 and 100 in Fig.\ref{att_rep_plot}. The SPS reaches a constant value after the interaction potential goes to zero for values of $r \geq 2.5 fm$. With the increase in \textit{k} value for $r > 1.5$, the SPS decreases, which is in accordance with the phase equation \ref{PFMeqn}. 

Also, with an increase in \textit{k} value for larger distances, the incident projectile tends to move away from the attractive core, resulting in a decrease in SPS with \textit{r}. Step like behaviour of phase shift in Fig.\ref{att_rep_plot} for smaller k indicates that bound states (here one bound state for r $<$ 1.5 and no bound state for r $<$ 0.15) are present in accordance with Levinson's theorem \cite{Levinson}, which states that at \textit{k = 0}, $\delta=n\pi$, where `n' denotes the number of bound states. This simply states that as the collision energy tends to zero (k $\rightarrow$ 0), the phase shift tends to $n\pi$, for bound states.

For similar attractive potential, with $\ell=2$, no bound state exists because of the centrifugal barrier, and it can be seen to be pierced by more energetic particles. The more energetic the scattering particles (\textit{for smaller distances shown in inset}), the more they penetrate into the potential region, and therefore, the larger the scattering phase shift as shown in the inset of Fig. \ref{fig:att_pot_l2}. On moving away from the attractive core, there is very weak potential, almost approaching zero, hence SPS goes down.
\begin{figure}[h!]
\centering
{\includegraphics[scale=0.42]{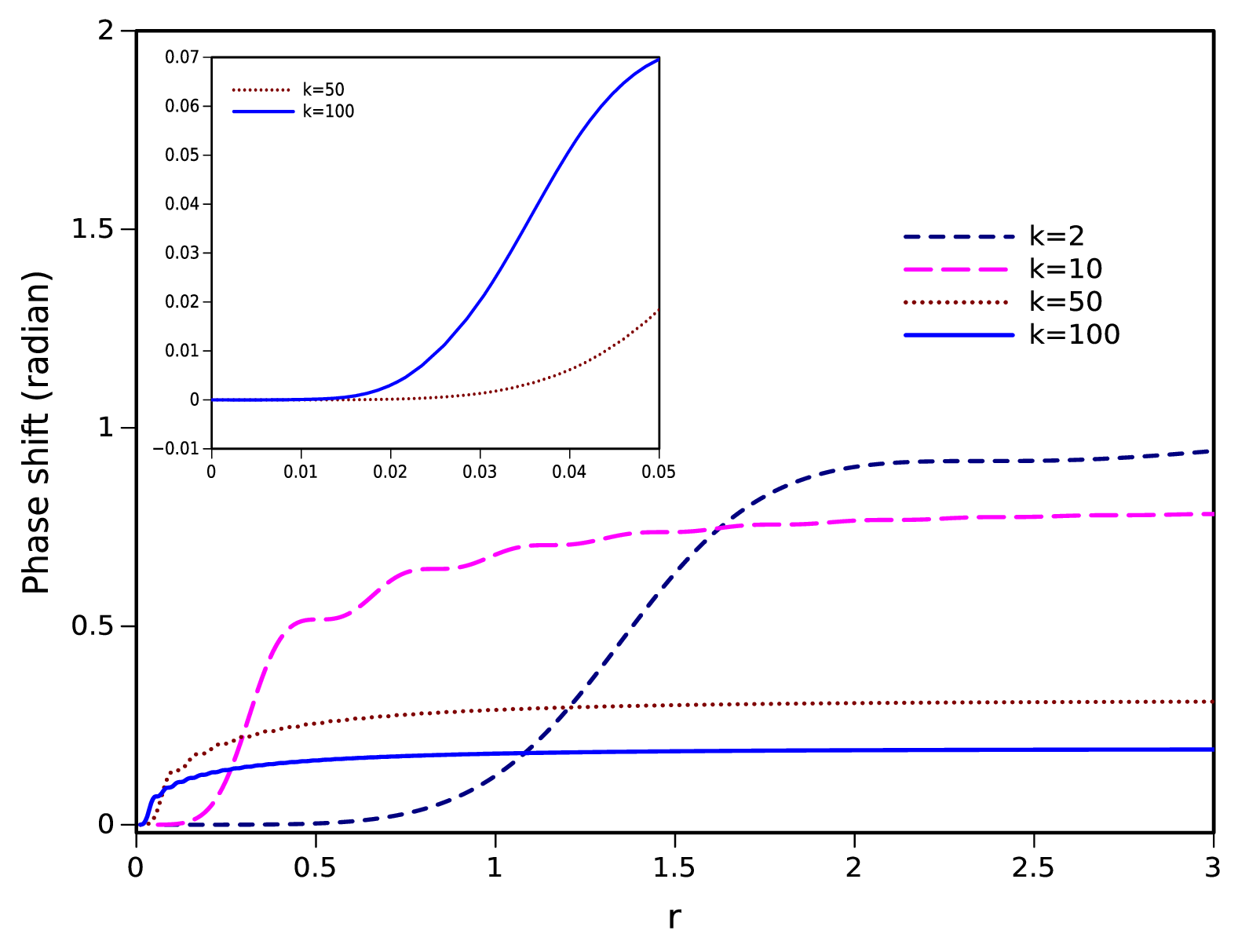}}
\caption{Phase shift variation with \textit{r} for similar attractive Yukawa potential but for $\ell=2$. Inset shows variation of phase shift for smaller distances.}
\label{fig:att_pot_l2}
\end{figure} 
\item \underline{Repulsive Yukawa:} $V(r)=(20/r)e^{-2r}$ for $\ell=0$:
The potential is having similar form except that it is having shorter range and greater strength. In case of repulsive potential, the phase shifts are negative which is in accordance with main VPA equation. In this case, for large, \textit{k} the particles face the repulsive core of the potential (r $\approx$ 0.03) and hence phase gets shifted by large angles. This is shown in inset of Fig. \ref{fig:rep_pot}. The accumulation of phase shift begins for smaller value of \textit{k} and larger value of \textit{r}. 
\begin{figure}[h!]
\centering
{\includegraphics[scale=0.42]{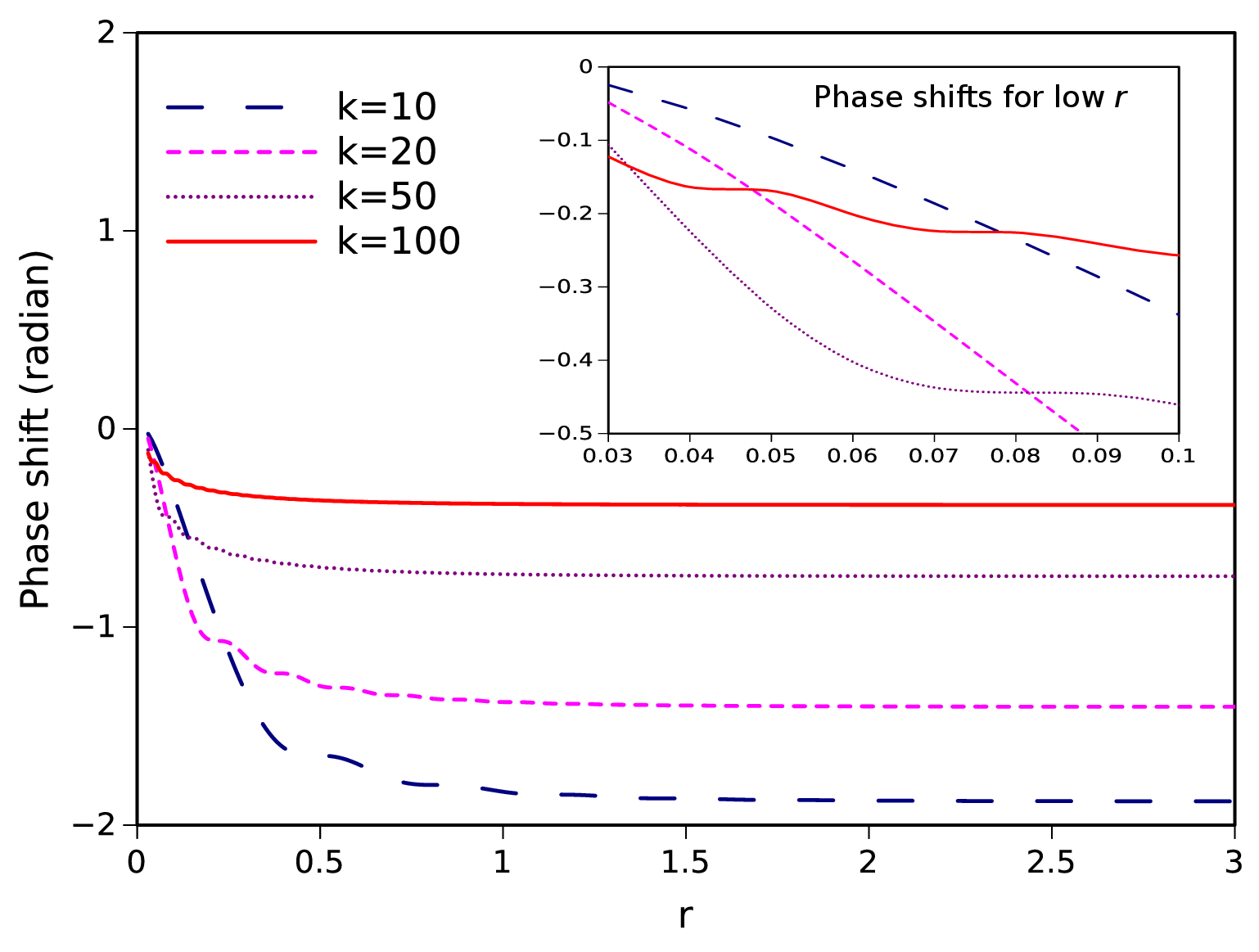}}
\caption{$\ell=0$ phase shifts at different \textit{k} values for Repulsive Yukawa potential.}
\label{fig:rep_pot}
\end{figure} 
\item \underline{Combination of attractive and repulsive Yukawa:} $V(r)=-(10/r)e^{-r}+(20/r)e^{-2r}$ for $\ell$ = 0, 2, 4 and 6. The Scilab-Xcos scheme for phase shift variation is shown in Fig.\ref{effect}. Fig.\ref{fig:both_pot} depicts some intricate phase shift plots when both positive and repulsive potential are present. For $\ell=0$ and k = 5, particles face the repulsive core of the potential, while in comparison, the k = 1 (less energy) particles, which face the attractive part of the potential. As a result, the phase shift is negative for (k = 5, $\ell = 0$) and increasing for (k = 1, $\ell = 0$). Because the particles may have encountered the repulsive core at some point in time, a small portion of the phase shift for (k = 1, $\ell = 0$) is negative. We can observe in the same (Fig.\ref{fig:both_pot}) that the phase shift changes from negative to positive for higher values of angular momentum because, as $\ell$ increases, the particle is forced away from the repulsive core toward the attractive core of the potential.
\begin{figure}[h!]
    \centering
    \hspace*{-2cm}
      \includegraphics[scale=0.45]{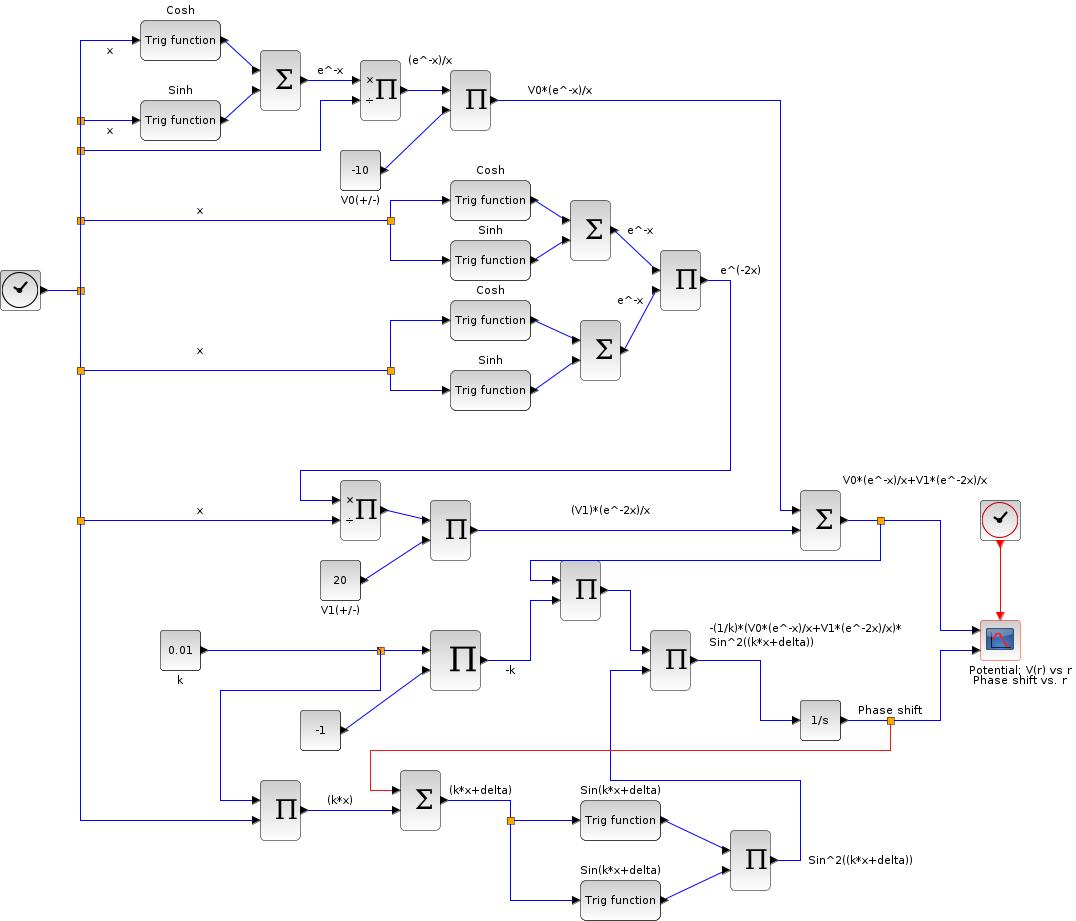}
    \caption{Scilab-Xcos scheme for phase shift variation for combination of attractive and repulsive potential ( $V(r)=-(10/r)e^{-r}+(20/r)e^{-2r}$) i.e.,  w.r.t distance.}
    \label{effect}
\end{figure}
\begin{figure}
\centering
\hspace*{-1.5cm}
{\includegraphics[scale=0.5]{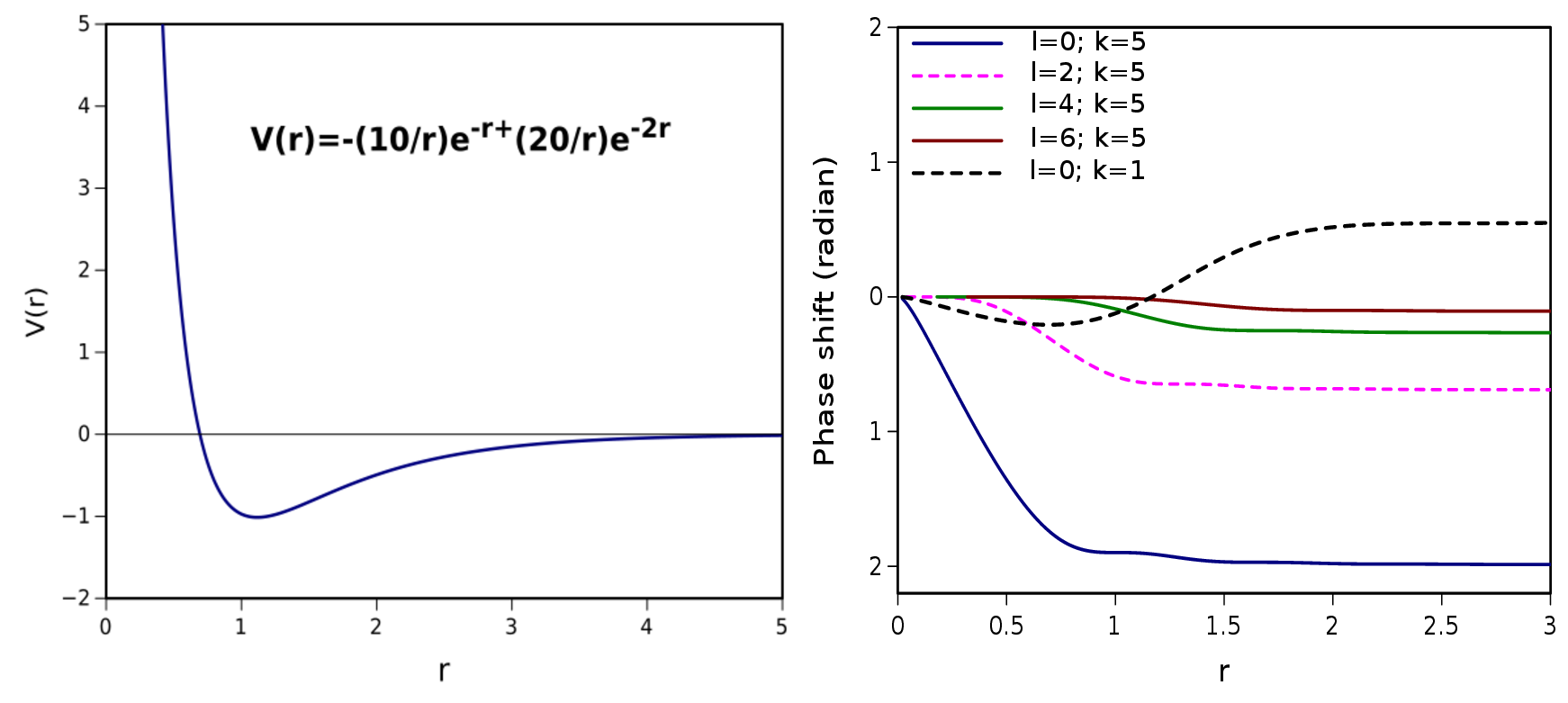}}
\caption{$\ell=0$ phase shift variation with combination of attractive and repulsive Yukawa potential $V(r)=-10/re^{-r}+20/re^{-2r}$.}
\label{fig:both_pot}
\end{figure}
\noindent Fig.\ref{fig:inone} depicts a phase variation within the potential. For a large k value, i.e., k = 10, the phase shift is negative because of the presence repulsive wall. For lower k values, i.e., k = 0.02 and 0.05, we have the following observations:
\begin{figure}
\centering
{\includegraphics[scale=0.75]{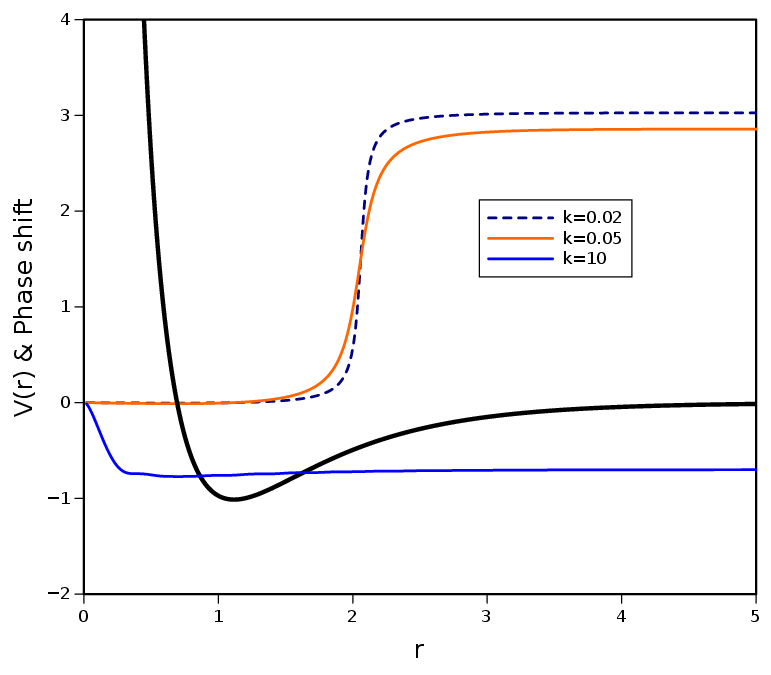}}
\caption{Variation of $\ell=0$ phase shift with combination of attractive and repulsive Yukawa potential $V(r)=-10/re^{-r}+20/re^{-2r}$. Variation of potential is shown in black solid line.}
\label{fig:inone}
\end{figure} 
\begin{itemize}
\item For r $<$ 0.7 $\rightarrow \delta$=-ve
\item For r $>$ 0.7 $\rightarrow \delta$=+ve
\end{itemize}
In summary, for large values of \textit{k}, the phase shift gets accumulated for smaller values of \textit{r} while for small values of \textit{k}, the phase shift gets accumulated for large values of \textit{r}. All discussed qualitative observations are in accordance to the main VPA equation. Also for attractive potential (V(r)$<$0), the phase shift is positive and for repulsive potential (V(r)$>$0), the phase shift is negative. This property is responsible  for the existance of repulsive core in strong intercations between two protons.    
\end{itemize} 
\section{Conclusion}
Xcos is a student-friendly toolbox of Scilab. In this paper, we demonstrate that a Riccati-type nonlinear differential equation, which is derived from TISE, can be easily solved using the simple blocks available in Xcos 
(GUI environment). In addittion to Kronig-Penney type periodic potential we have utilized three different Yukawa forms of spherically symmetric potentials: (i) attractive Yukawa, (ii) repulsive Yukawa, and (iii) attractive and repulsive Yukawa potentials. Phase shift vs. r has been obtained for the stated potentials, illustrating how different potentials influence the scattering outputs. For the first time, an Xcos environment has been created to solve the Riccati type non-linear differential equation, which students can readily implement for various phenomenological potentials available in nuclear, atomic, and molecular fields. The paper can be easily extended to obtain the variation of phase shift with energy by simply noting the phase shift value obtained at the asymptotic region (V(r)=0) of the used potential.

\end{document}